# Midiendo la calidad de la información gestionada: algunas reflexiones conceptuales-metodológicas

**Carlos Luis González-Valiente**
Departamento de Informática y Gestión de la Información, Grupo Empresarial de la Industria Sidero Mecánica (GESIME) - Cuba

**ANÁLISIS**

## Resumen

El estudio, basado en un análisis documental clásico, reflexiona sobre algunas directrices conceptuales-metodológicas relativas a la medición de la calidad de la información (CI) en el marco de la gestión de información (GI) en las organizaciones. Es descrito el proceso de GI y la importancia de la aplicación de principios de calidad en éste. Se exponen las cuatro dimensiones de la CI según una integración de los indicadores esenciales que caracterizan a los contenidos informacionales. Son definidas cada una de las fases que componen el diseño de una metodología para evaluar la información. Se indican las implicaciones que tiene este ejercicio para los profesionales del campo informacional..

## Palabras clave

*Calidad de información, Gestión de información, Dimensiones de calidad de información, Metodología de calidad de información*

## Measuring the quality of information managed: some conceptual and methodological reflexions

## Abstract

The study, based on a documental classic analysis, presents conceptual and methodological guidelines concerning the design of methodologies that help to measure the quality of information that is managed in organizations. It is described the process of information management and the importance of implementing quality principles in it. There are exposed the four dimensions of information quality as part of an indicators integration which characterize the informational contents. There are defined each of the phases in the methodological design to evaluate the information. There also are indicated the implications of this activity for information professionals.

## 1. Introducción

Investigaciones enfocadas en las temáticas de CI se concentran mayormente en cuestiones de calidad desde la perspectiva de los sistemas, productos o servicios. Esto lo corrobora una búsqueda exploratoria en la base de datos ScienceDirect (1), en la cual bajo la ecuación de búsqueda Information Quality en el título del conjunto de revistas indizadas, fueron determinados como tópicos más abordados: internet, sitio web, calidad de servicio, sistema de información y tecnología de información.

Es importante comprender que la determinación de la CI es una actividad implícita en procesos que son puramente informacionales. En el contexto específico de la GI en las organizaciones, ésta como actividad informativa bien integradora, la CI es crucial porque las decisiones dependen de la buena formulación que los profesionales de la información sean capaces de ejecutar (Ballou & Tayi, 1999).

Este artículo pretende reflexionar sobre algunas directrices conceptuales-metodológicas relativas a la medición de la CI en el marco de la GI en el dominio organizacional. Mediante un análisis de algunas investigaciones relevantes que han abordado esta temática se delimitarán los elementos básicos en el diseño de enfoques metodológicos sobre CI. Por último, se enunciarán las implicaciones que tiene el ejercicio de esta actividad para los profesionales del campo informacional, con la finalidad de destacar pistas esenciales para el mejoramiento de sus tareas de trabajo.

## 2. La información y su proceso de gestión

Antes de comenzar deconstruyendo cada una de las categorías de análisis de este estudio se precisará conceptualmente qué es la información, debido a las diferentes interpretaciones que le es conferida. De forma general se asocia información a dato y a conocimiento; y aunque esta tríada presenta una gran interrelación, la naturaleza de sus individuales significados es muy cuestionable (Zins, 2007). Disparmente, la literatura sobre gestión lo vincula más a conocimiento (Bates, 2006), mientras que los abordajes específicos sobre CI lo relacionan al dato, debido al vínculo directo de esta temática con los sistemas tecnológicos. El interés de este artículo es explorarla desde la perspectiva en la que ésta puede ser gestionada, por ello se paraleliza información y dato como elementos que, bajo profundos parámetros de análisis, llenan vacíos de conocimiento que permiten llevar a cabo o redirigir acciones concretas.

La GI es considerada por Ponjuán (2011) como la actividad o función estratégica que concreta las políticas de información de las instituciones. Más estructuralmente, Choo (1997) la operacionaliza en una red de seis procesos interrelacionados: identificación de necesidades de información, adquisición de información, organización y almacenamiento de información, desarrollo de productos y servicios de información, y diseminación y uso de la información. Ellos son delimitados a partir de la premisa de que las organizaciones se comportan como un sistema abierto que transforma la información en conocimientos, procesos y estructuras que generan, a su vez, nuevos productos y servicios (Choo, 1996). En dicha transformación están implícitas actividades claves como la determinación de la calidad de los contenidos informacionales, considerado por Ponjuán (2002) como uno de los elementos que integra el alcance de la GI en el entorno organizacional.

La materialización de la información como un recurso crítico para la toma de decisiones y la gerencia (Alwis de & Higgings, 2001) presupone que la calidad de la planeación estratégica dependa de la calidad de los recursos informativos adquiridos tras el análisis de las numerosas fuentes disponibles (Auster & Choo, 1994). Sin embargo, es importante señalar algo que Ponjuán (2011) precisa conceptualmente cuando valora los modelos de GI de Butcher y Rowley (1998), Paéz Urdaneta (1992), Choo (1992) y Ponjuán (2000). No debe confundirse este concepto cuando se aborde "lo referido al acceso y uso de la información o a su tratamiento, procesamiento y organización" (p.16). Cada una de estas tareas puede darse independientemente una de otra y caer perfectamente bajo los marcos conceptuales del comportamiento informacional (2) que las personas asumen a la hora de manejar la información, lo cual no implica que se esté hablando en términos puramente de GI. Ponjuán (2011) distingue que esta actividad debe concebirse y manejarse como una función estratégica institucional y no como proceso operativo que asegura el tratamiento de la información y la prestación de servicios.

## 3. Dimensiones de la calidad de la información

Peter (2011) afirma que la información debe poseer calidad para que sea considerada como información. Precisamente, muchos autores consideran que la información posee calidad cuando ésta es completamente útil en la toma de decisiones o la solución de algún problema específico. Por lo que si el proceso que conlleva a la determinación de la calidad informativa falla, la decisión tomada o el problema a solucionar no serán factibles, ya que un factor es condicionante preciso del otro. Esto tiene lugar, en gran medida, por la marcada subjetividad subyacente en cada una de las estructuras mentales de quienes encaminan esta actividad. Aquí los patrones, modelos y estructuras cognitivas definen, debido a que las personas no solo crean construcciones subjetivas de su experiencia, sino que esas construcciones de la información también tienen una existencia objetiva en el sistema nervioso y son externalizadas en acciones específicas (Choo, 1996).

Van Birgelen, De Ruyter y Wetzels consideran, similarmente, que la utilización de la información de calidad está determinada por el nivel de satisfacción del proceso y resultado de un proyecto de investigación específico (2001). El gráfico 1 representa a dicha satisfacción como el componente de núcleo de cualquier proyecto.

**Gráfico 1 – Marco conceptual del uso de la información relacionada com la calidad**

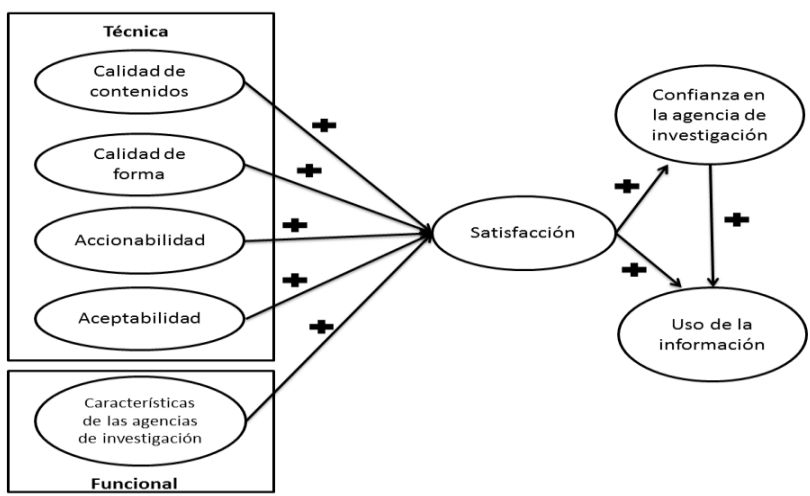

Fuente: Van Birgelen, De Ruyter & Wetzels, 2001.

La marcada subjetividad implícita en un proceso de CI ha impulsado la multiplicidad de variables que hacen factible su determinación y evaluación. Su variabilidad viene dada mayormente por la relación objeto de información-contexto de información, es decir, una proporcionalidad entre contenido-forma y el contexto situacional. En ese contexto es donde entrarían agentes como las capacidades de las personas, los requerimientos de la organización, los procesos asociados, etc. Para Stvilia (2007) el contexto posee dos componentes básicos: la cultura (lenguaje, normas) y las estructuras sociotécnicas (incluyendo relaciones económicas y estándares). Del objeto de información es determinante la particularidad o atributo específico que es importante medir.

Las dimensiones son uno de los principales componentes de la CI, y éstas son agrupadas por Lee et al. (2008) en cuatro categorías: intrínseca, contextual, representacional y acceso (Véase tabla 1). Estas categorías han sido previamente definidas por Strong (1997), quien las considera como las que mayor calidad le atribuyen a la información. Para estos autores la dimensión intrínseca concibe a la información como algo que tiene calidad por derecho propio. Mientras que la contextual destaca que los requerimientos de calidad se dan en un contexto dado y que la información tiene que ser entonces relevante, oportuna, completa y apropiada en términos de cantidad; así como de agregación de valor. Por último, las dimensiones representacional y de acceso enfatizan la importancia de los sistemas de cómputo para el almacenamiento y el acceso de la información.

Tabla 1. La mirada académica de la calidad de la información

|  | Intrínseca | Contextual | Representacional | Acceso |
|---|---|---|---|---|
| **Wang and Strong (1996)** | Precisión, credibilidad, reputación, objetividad | Valor añadido, pertinencia, completa, oportunidad, cantidad apropiada | Entendible, interpretable, representación concisa, representación consistente | Accesibilidad, facilidad de operaciones, seguridad |
| **Zmud (1978)** | Precisa, factual | Cantidad, confiable / tiempo | Disponible, leíble, razonable |  |
| **Jarke and Vassiliou (1997)** | Creíble, veracidad, credibilidad, coherencia, completa | Relevancia, uso, oportuna, circulación de la fuente, circulación de almacén de datos, no volátil | Interpretabilidad, sintaxis, control de versiones, semántica, renombrada, origen | Accesibilidad, disponibilidad del sistema, disponibilidad de transacción, privilegios |
| **Delone and McLean (1992)** | Exactitud, precisión, fiabilidad, libertad de prejuicios | Importancia, relevancia, utilidad, informativa, contenido, suficiencia, completa, vigente, oportuna | Entendible, leíble, claridad, formato, apariencia, conciso, originalidad, comparable | Usabilidad, cuantificable, conveniencia de acceso |
| **Goodhue (1995)** | Precisión, fiabilidad | Vigencia, nivel de detalles | Compatibilidad, significado, precisión, falta de confusión | Accesibilidad, asistencia, uso fácil, localizable |
| **Ballou and Pazer (1985)** | Precisión, consistencia | Completa, oportuna |  |  |
| **Wand and Wang (1996)** | Correcta, no ambigua | Completa | Significación |  |

Fuente: Lee et al., 2008.

En estas dimensiones, nuevas propuestas han sido expuestas en la literatura científica, como es el caso de Ruby y Vashchilko (2012), quienes sugieren la veracidad/decepción (3), como atributo que debe incluirse en la categoría intrínseca. Cada dimensión funciona como un marco conceptual y de referencia para la modelación de la cualificación informativa. Pues aquí se integran las variables apropiadas que fungen como objeto para la ejecución de las evaluaciones apropiadas. Según Stvilia et al. (2007) "el marco de evaluación de la CI como estructura multidimensional consiste en conceptos, relaciones, clasificaciones y metodologías generales que pudieran servir como un recurso o guía para desarrollar modelos de medición de CI en un contexto específico" (p.1722).

## 4. Procedimiento para la determinación y medición de la calidad informativa

Múltiples perspectivas son tomadas para elaborar enfoques metodológicos que ayudan a determinar la calidad de los contenidos informacionales que diariamente se gestionan. Entre las nueve perspectivas existentes se destacan las enunciadas por Betini et al. (2009):

1) Las fases y los pasos que componen la metodología
2) Las estrategias y los métodos para medir o evaluar la información
3) Las dimensiones y las metrías adoptadas
4) Los tipos de costos asociados
5) Los tipos de datos que son considerados
6) Los tipos de sistemas de información que usan, modifican y manejan los datos
7) Las organizaciones incluidas en los procesos que crean o eliminan los datos

8) Los procesos que crean o eliminan los datos que tributan a la producción de servicios
9) Los servicios que son producidos por los procesos considerados en la metodología

La multiplicidad de estas metodologías varía según el contexto de aplicación, los procesos asociados y los resultados a los que dichos procesos tributan. Capiello y Pernici (2006) y Batini *et al.* (2009) coinciden en que son cuatro las fases que la organización debe comprender para determinar dicha calidad; éstas son: la definición de la calidad de los datos, la medición, el análisis y el mejoramiento. En respuesta a los planteamientos de estos autores, en este estudio solo se tendrán en cuenta las tres primeras perspectivas enunciadas anteriormente.

### 4.1. Definición de la calidad de la información

En esta primera se recolecta toda la información y se definen los indicadores a medir, teniendo en cuenta cada una de las cuatro categorías que componen las dimensiones de calidad expuestas por Lee et al (2008). Es necesario, constar de una exploración o análisis previo del entorno en el cual la información se encuentra o transita. Puede que sea, por ejemplo, un sitio web o una red social, un sistema de información implementado, un sistema de almacenamiento de datos que la organización posea o una base de datos.

Para ello es necesario identificar el proceso organizacional en el cual la medición se enmarca, en correspondencia con los requerimientos de información de los usuarios (empresa, departamento, equipo de trabajo especializado, especialista individual). Strong (1997) considera que la CI no puede ser medida independientemente de las personas que la usan, donde la mirada, además de dirigirse a los usuarios o consumidores, debe dirigirse también a los productores y consumidores de información.

### 4.2. Medición de la calidad de la información

Los métodos para medir o evaluar la calidad de la información son usados de acuerdo a los resultados que se deseen obtener. Como mayormente esta es una actividad que está condicionada por la perspectiva subjetiva de los individuos (percepción, razonamiento, interpretación), es factible la combinación de técnicas cuantitativas (análisis numéricos y estadísticos) y cualitativas (entrevistas, cuestionarios). Por ejemplo, Lee et al. (2008) diseñaron un cuestionario y aplicaron un muestreo para medir la CI en una organización; para ello delimitaron ítems que respondían a cada una de las categorías de las dimensiones antes mencionadas. A cada uno de estos ítems les fue asignada una escala de valor con la cual se determinó el comportamiento de cada variable dentro de las dimensiones. Por otro lado, Eppler y Muenzenmayer (2002) han expuesto métodos para el entorno específico de la web, entre los cuales destacan los siguientes:

a) El monitoreo de rendimiento

b) El analizador de sitio

c) El analizador de tráfico

d) La minería web

e) La retroalimentación del usuario

En los argumentos de su investigación afirman que de estas cinco tipologías existen software disponibles, específicamente para el tipo a), b) y c). Mientras que para d) y e) hay herramientas más poderosas pero muy costosas. Ahora, lo que sí varía en estas cinco tipologías es la presencia de atributos de la información sobre los cuales no es posible aplicar cualquier tipo de variable. Además, que dicha tipología responde más al medio (forma) que al modo (contenido) de la información; aunque ambas cuestiones son importantes porque resulta difícil desliar un elemento del otro. De ahí la relevancia de identificar previamente cuáles son los parámetros de calidad que se desean evaluar, según lo expuesto en la tabla 1. Es oportuno destacar que sobre esta temática muchas de los estudios emergidos de la literatura exponen evaluaciones a partir de ecuaciones algebraicas u otra clase de análisis desde la perspectiva de los sistemas tecnológicos y procesos de software (ej. Su & Yi, 2008; Caballero, Caro & Piattini, 2008).

### 4.3. Análisis del resultado de la medición

Para la fase de análisis hay autores que abogan por la categorización de los elementos obtenidos (ej. Fink-Shamit & Bar-Ilan, 2008); en este caso referido a los atributos incluidos en cada dimensión (Véase tabla 1). A cada grupo de elementos se le puede asignar un valor tanto numérico como descriptivo. En esta tarea pueden darse niveles de participación múltiple o individual para analizar los resultados obtenidos, en donde los especialistas emiten su criterio para identificar los patrones emergentes durante el proceso evaluativo. Un modelo simple para ello puede ser el propuesto en la tabla 2, en donde convergen externamente el proceso específico de GI y el requerimiento implantado por la organización. Estos requerimientos aquí se traducen como algo circunstancial; y no son más que un problema dado o una necesidad expresa sobre la cual la GI se está enfocando. Dicha GI puede orientarse tanto a un proceso organizacional (investigación de mercado, análisis del entorno, inversiones, transacciones, etc.), como a un servicio o producto, según la fase de desarrollo o análisis de éstos. Sobre ellos serán distinguidos y descompuestos los elementos estructurales que lo integran como lo ejemplifica la tabla 2. En la Dimensión de la CI se expondrán los atributos a los cuales se les añadirá un valor deseado y una breve descripción del resultado de la evaluación.

**Tabla 2. Modelo para medir la calidad de la información**

| Requerimiento organizacional (Proceso, producto o servicio) | Dimensión de la CI | Medición/Valor | Descripción |
|---|---|---|---|
| | | | |
| | | | |

Fuente: Elaboración propia.

### 4.4. Mejoramiento de la calidad de la información

El mejoramiento de la información es una de las etapas que se tiene en cuenta cuando la finalidad del proceso de CI va dirigida a un producto, servicio o sistema de información específico. Ello incide en que las prácticas de gestión sean replanteadas y por lo tanto se desecha o acepta la información utilizada. Caballero *et al.* (2008) afirman que el primer paso es realizar una evaluación del escenario en el cual la mejora se realizará (negocios, web, software, sistemas de gestión), aunque actualmente no existen muchos modelos de referencia que permitan demostrar cuán optimo un escenario puede ser. Batini *et al* (2009) destacan que en esta etapa se definen las medidas, estrategias y técnicas para alcanzar los nuevos objetivos de calidad de la información.

### 5. Implicaciones profesionales

Si se ha estado hablando de la necesidad de aplicar enfoques de CI en el proceso de GI, es clave identificar como nexo la perspectiva humana, la de los profesionales de la información. Ballou y Tayi (1999) confirman que los gestores de la información, en conjunto con los usuarios de ésta, deben pensar sistemáticamente en función de lo que es necesario, vital y deseado. Además de que estas iniciativas deben estar guiadas de acuerdo a las ideas, políticas y tendencias de CI que están presentes en la organización. Esto encuadra perfectamente con lo enunciado por Ponjuán (2011), cuando afirma que la GI se materializa en términos de políticas de información específicas. Sin embargo desde una perspectiva inversa, o como una forma de contribución, Caballero *et al* (2008) indican que las responsabilidades de estos especialistas van encaminadas a determinar las políticas de CI según los requerimientos organizacionales, de manera tal que se puedan identificar de forma certera las técnicas, herramientas y procedimientos para evaluar el recurso información. Siempre teniendo como base el contexto en el cual ésta transita y se usa. Pues ello contribuiría a definir modelos conceptuales y estratégicos según los procesos específicos a los que la GI se orienta.

**Consideraciones finales**

La gestión constante de información es una práctica que debe estar condicionada por la aplicación de parámetros de calidad que permitan que la información suministrada, compartida y usada posea los requisitos óptimos. Actualmente son muchas las metodologías que existen, las cuales difieren en dependencia del contexto organizacional que se aplica. Dicho contexto está básicamente definido por los procesos y las personas implícitas en ellos. Esto conduce a la construcción de modelos de calidad de información que responden a entornos específicos, los cuales deben tener como base conceptual cada una de las dimensiones de calidad informativa (*intrínseca*, *contextual*, *representacional* y *accesible*) declaradas. Los parámetros metodológicos que aquí se resumen están mayormente dirigidos al desarrollo de sistemas; pero aplica de igual forma a todo tipo de prácticas de GI, porque esta actividad debe atender a los componentes del mismo y a su vez optimizarlo de manera que se puedan ejecutar efectivamente las funciones de gestión (Ponjuán, 2011).

Es bueno considerar que cualquier práctica informativa requiere de garantías de calidad, lo cual presupone que los profesionales de la información se responsabilicen por los procesos de producción informativa a través de cada una de las etapas del ciclo de gestión de información a los que Choo (1996) hace alusión. Pues estos especialistas fungen como filtros humanos que, mediante conocimientos y habilidades oportunas, inciden en la calidad dada en la relación establecida entre el usuario del sistema y la información provista.

**Notas**

1. *ScienceDirect*, en su sitio web, se presenta como una de las bases de datos científica de la compañía Elsevier, la cual ofrece acceso a artículos de revistas y capítulos de libros que han sido previamente arbitrados. Posee además una infraestructura adecuada para la ejecución de búsquedas que son oportunas para el análisis bibliométrico de los contenidos (http://info.sciencedirect.com/about).

2. El comportamiento informacional estudia "cómo las personas necesitan, buscan, comparten y usan la información en diferentes contextos, incluyendo el lugar de trabajo y la vida diaria" (Pettigrew, Fidel & Bruce, 2001, p.45).

3. Para mayor profundidad de lo que se propone desde este componente en la CI, consúltese a Ruby y Vashchilko (2012).

## Bibliografía

## Datos del autor

### Carlos Luis González Valiente
Licenciado en Ciencias de la Información por la Universidad de la Habana. Su ámbito de investigación son la alfabetización informacional y la mercadotecnia en el campo de las disciplinas informativas. Es especialista en información del Departamento Independiente de Informática y Gestión de la Información, del Grupo Empresarial de la Industria Sidero Mecánica de Cuba.
carlos.valiente@fcom.uh.cu